\documentclass{ws-procs975x65}
\begin{document}

\title{The upgrade of GEO\,600}

\author{H. L\"uck$^*$ for the LIGO Scientific Collaboration}

\address{Max-Planck-Institut f\"{u}r Gravitationsphysik (Albert-Einstein-Institut) and Leibniz Universit\"{a}t Hannover,
Callinstr. 38, 30167 Hannover, Germany\\
$^*$E-mail: harald.lueck@aei.mpg.de}

\begin{abstract}
This article gives a short overview of the status of the British/German gravitational wave detector GEO\,600 and the upgrades planned within 2010 and 2011
\end{abstract}


\bodymatter

\section{Status Quo}\label{status}
The German/British interferometric gravitational wave (GW) detector GEO\,600 has taken data for almost 3.5 years from the beginning of 2006 to July 2009, partly together with the other large GW detectors LIGO and Virgo in the S6/VSR1 data run (Jan. 2006 to Oct. 2007) and partly together with the 2\,km H2 LIGO detector during 'Astrowatch' (Nov. 2007 to July 2009) in the upgrading process of the Virgo and LIGO detectors. All these detectors are cooperating in the LSC/Virgo network and jointly analyze the collected data, whenever simultaneously taken data are available at comparable sensitivities. For an overview of the worldwide GW network refer to the article of David Reitze in this issue.\par
\begin{figure}[h]
\begin{minipage}{0.5\textwidth}
\centering
\includegraphics[width=\textwidth]{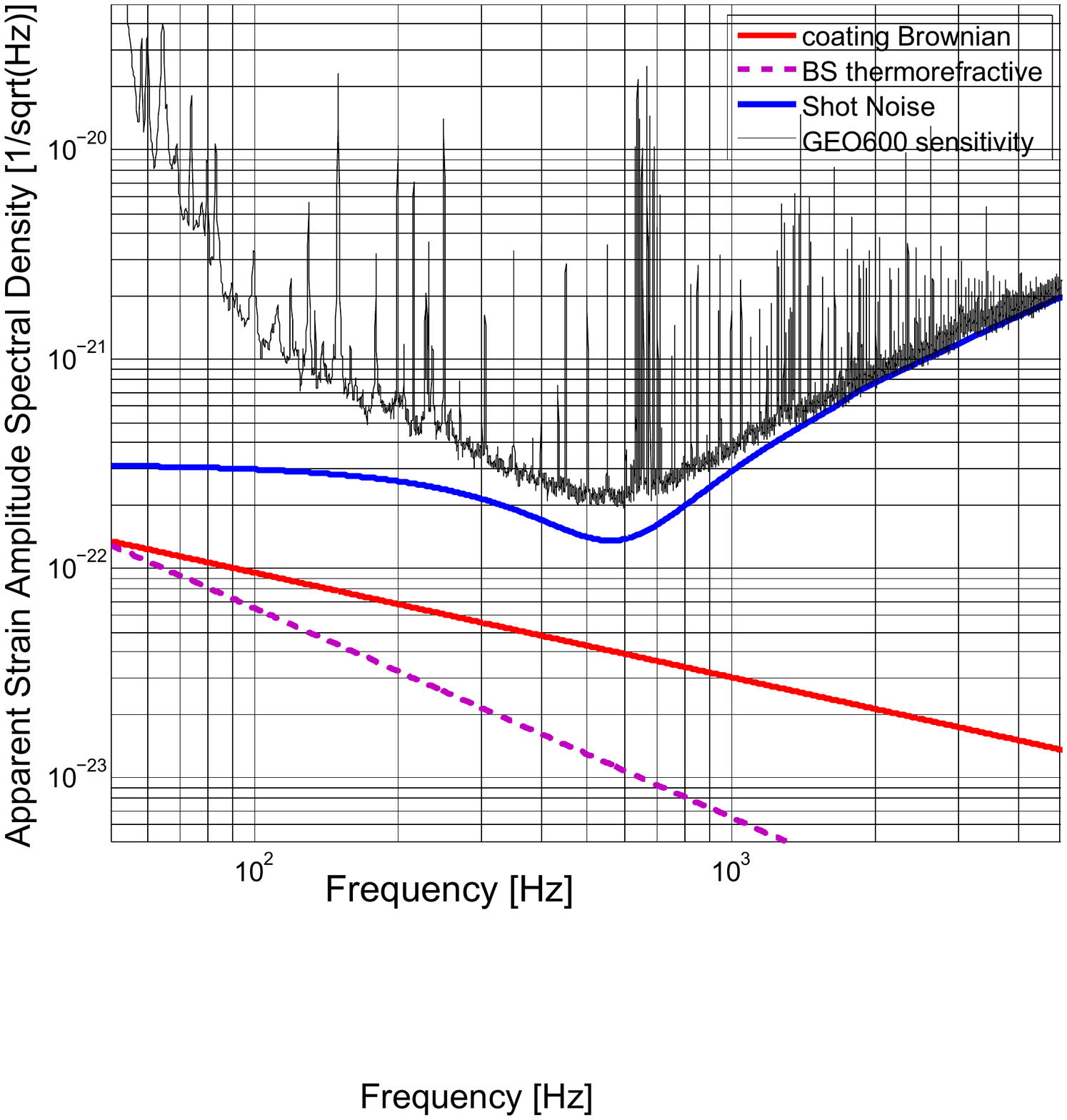}
\caption{Sensitivity of GEO\,600 as of August 2009}
\label{GEOSensitivity} 
\end{minipage}
\begin{minipage}{0.04\textwidth}
\hspace{0.5cm}
\end {minipage}
\begin{minipage}{0.44\textwidth}
Above 500\,Hz the sensitivity of GEO\,600 is currently limited by photon shot noise 
while technical noise sources dominate at lower frequencies. Coating Brownian thermal noise and thermo-refractive noise of the beam splitter are the dominant 'fundamental' noise sources at low frequencies. With the following changes the upgrades of GEO\,600 will therefore aim at the frequency range above 500\,Hz.
\end{minipage}
\end{figure}

 
Being the smallest of the detectors in the network, GEO\,600 implemented several 'advanced' techniques to reach a comparable sensitivity\cite{GEO600Grote2008}. Dual Recycling\cite{DualRecExpStrain1991}, monolithic suspensions\cite{ViolinAndMSGossler2004} and electro-static actuators\cite{GEOESDChargingHewitson2007} are novel techniques that are being used in the current GEO\,600 configuration but are only foreseen for the advanced stages of LIGO and Virgo. Throughout the data taking period GEO\,600 used 6\,W of laser power incident to the Mode Cleaners\cite{GEOMCgossler2003}, i.e. two subsequent optical resonators of 8\,m round-trip length 
, which remove optical higher order modes form the laser beam and thereby reduce angular beam fluctuations. Due to scattering losses in these Mode Cleaners about 3.2\,W of laser power can be used at the interferometer. The GW signal and with it the longitudinal and angular control signals for the main interferometer are generated with the Schnupp modulation technique\cite{Schnupp1988}, which involves phase modulating the laser beam before it enters the interferometer and demodulating the output beam at the same frequency. GEO\,600 uses both Power- and Signal-Recycling to enhance the usable laser power inside the interferometer by factor of about 1000 and to enhance the signals at the output port in a band of about 700\,Hz around the resonance frequency of the Signal Recycling cavity. This resonance frequency can be tuned by the exact position of the Signal Recycling mirror ($\approx$ 4\,Hz/pm) and was set to 530\,Hz during the data taking runs. With this set-up a peak sensitivity of 2E-22 /sqrt(Hz) has been reached (see figure\ref{GEOSensitivity}).

\vspace{-0.3cm}
\paragraph {DC readout}
Schnupp modulation as mentioned above increases the detected shot noise above the fundamental limit by mixing shot noise from twice the modulation frequency into the detection band\cite{Buonanno2003}. By direct detection of the light power at the output port the shot noise can be lowered. GEO\,600, LIGO and Virgo will use this so called DC read-out technique\cite{HildDCReadout2009}.
\vspace{-0.3cm}
\paragraph {Output Mode Cleaner}
Due to deviations of the mirror surfaces from an ideal sphere higher order optical modes of the Signal Recycling cavity exit to the output port of GEO\,600. This light adds shot noise but does not contain GW signals. By removing this light with an additional optical resonator in the output port, the Output Mode Cleaner (OMC\cite{DegallaixAmaldi2009}), the signal to noise ratio can be improved. GEO\,600 will use a 4-mirror OMC with a round trip length of 66\,cm and a finesse of $\approx$150.
\vspace{-0.3cm}
\paragraph {Squeezing}
The injection of light squeezed in the phase quadrature into the output port can lower the shot noise of an interferometric GW detector\cite{OpticalSpringHarms2003}. Using a squeezing level of about 10\,dB and allowing for losses of about 15\% on the way from the squeezer via the signal recycling (SR) mirror, through the OMC to the photo diode will reduce the shot noise by a factor of two in strain amplitude spectral density, equivalent to a power increase of a factor of four.
\vspace{-0.3cm}
\paragraph{Tuned, broadband Signal Recycling}
To make use of squeezing over the full desired frequency range the squeezed light must undergo the same frequency dependent phase rotation as the light exiting the SR cavity. This is the case if the SR cavity is tuned to carrier resonance and both signal sidebands see the same resonance conditions. Changing the reflectivity of the SR mirror from 2\% to 10\% will widen the bandwidth of GEO\,600 to about 3.5\,kHz to improve the high frequency performance.
\vspace{-0.3cm}
\paragraph {Light power increase}
The signal strength can be increased by increasing the light power inside the interferometer. At the same time the light power on the photo detector will increase yielding higher shot noise. The signal increases linearly with the light power whereas the shot noise only rises with the square-root. Hence a net gain with the square root of the light power is achieved. In GEO\,600 the laser power used at the input to the mode cleaners will be increased from 6\,W to 35\,W.\par
With increasing light power, radiation pressure effects in the mode cleaners become increasingly problematic. The increased laser power can be compensated by lowering the finesse by a factor of 5, yielding the same intra-cavity light power as before. The reduced susceptibility to optical losses will raise the throughput by almost a factor of 2 and will increase the laser power inside the interferometer from about 3\,kW to 30\,kW. Thermal lensing from residual absorption in the beam splitter will be compensated by appropriately irradiating the beam splitter with infrared light from an incandescent source. The resonances of the triple pendulum mirror suspensions in GEO\,600 are dampened at the upper stage by sensing the motion with shadow sensors and feeding back to magnet-coil actuators. These shadow sensors operate with DC LEDs as a light source and can sense light scattered from the main interferometer beam. This light disturbs the control loops and leads to mirror misalignments. In order to decrease this coupling and avoid problems after the light power increase, the shadow sensors will be operated with amplitude modulated light.
\vspace{-0.3cm}
\section*{Acknowledgments}
The authors gratefully acknowledge the support of the United States
National Science Foundation for the construction and operation of the
LIGO Laboratory and the Science and Technology Facilities Council of the
United Kingdom, the Max-Planck-Society, and the State of
Niedersachsen/Germany for support of the construction and operation of
the GEO600 detector. The authors also gratefully acknowledge the support
of the research by these agencies and by the Australian Research Council,
the Council of Scientific and Industrial Research of India, the Istituto
Nazionale di Fisica Nucleare of Italy, the Spanish Ministerio de
Educaci\'on y Ciencia, the Conselleria d'Economia, Hisenda i Innovaci\'o of
the Govern de les Illes Balears, the Royal Society, the Scottish Funding 
Council, the Scottish Universities Physics Alliance, The National Aeronautics 
and Space Administration, the Carnegie Trust, the Leverhulme Trust, the David
and Lucile Packard Foundation, the Research Corporation, and the Alfred
P. Sloan Foundation. Also the authors gratefully acknowledge the support of the Deutsche Forschungsgemeinschaft through the Transregio SFB TR7.
\vspace{-0.6cm}
\bibliographystyle{ws-procs975x65}
\bibliography{MG12GEO}

\begin{thebibliography}{10}

\bibitem{GEO600Grote2008}
H.~Grote {\em et~al.}, {\em Classical and Quantum Gravity} {\bf 25}, p. 114043
  (9pp) (2008).

\bibitem{DualRecExpStrain1991}
K.~A. Strain and B.~J. Meers, {\em Phys. Rev. Lett.} {\bf 66}, 1391(Mar 1991).

\bibitem{ViolinAndMSGossler2004}
S.~Gossler {\em et~al.}, {\em Classical and Quantum Gravity} {\bf 21}, S923
  (2004).

\bibitem{GEOESDChargingHewitson2007}
M.~Hewitson {\em et~al.}, {\em Classical and Quantum Gravity} {\bf 24}, 6379
  (2007).

\bibitem{GEOMCgossler2003}
S.~Gossler {\em et~al.}, {\em Review of Scientific Instruments} {\bf 74}, 3787
  (2003).

\bibitem{Schnupp1988}
L.~Schnupp, Presentation at european collaboration meeting on interferometric
  detection of gravitational waves, (Sorrent, Italy, Oct 1988), (1988).

\bibitem{Buonanno2003}
A.~Buonanno, Y.~Chen and N.~Mavalvala, {\em Phys. Rev. D} {\bf 67}, p.
  122005(Jun 2003).

\bibitem{HildDCReadout2009}
S.~Hild {\em et~al.}, {\em Classical and Quantum Gravity} {\bf 26}, p. 055012
  (10pp) (2009).

\bibitem{DegallaixAmaldi2009}
J.~Degallaix, Commissioning of the tuned dc readout at geo600, (subm. to JPCS).

\bibitem{OpticalSpringHarms2003}
J.~Harms {\em et~al.}, {\em Phys. Rev. D} {\bf 68}, p. 042001(Aug 2003).

\end{thebibliography}

\end{document}